\documentclass[twocolumn,showpacs,showkeys,preprintnumbers,amsmath,amssymb,amsfonts,euscript,revsymb,prd]{revtex4}
\usepackage[dvips]{graphicx}
\usepackage{hyperref}
\usepackage{dcolumn}
\usepackage{bm}

\expandafter\ifx\csname package@font\endcsname\relax\else
 \expandafter\expandafter
 \expandafter\usepackage
 \expandafter\expandafter
 \expandafter{\csname package@font\endcsname}%
\fi

\begin{document}

\title{On quantum cosmology as field theory of bosonic string mass groundstate}

\author{L.A. Glinka}
\email{laglinka@gmail.com}
\affiliation{Bogoliubov Laboratory of Theoretical Physics,\\\mbox{Joint Institute for Nuclear Research, 141980 Dubna, Russia}}
\date{\today}

\begin{abstract}
The Quantum Cosmology can be understand as the theory of an one object that is the Universe described in terms of fundamental mass groundstate of the free bosonic string, that is a tachyon - a hypothetical particle with negative mass square, which has linear velocity more than the velocity of light $c$. From this fact it is clear that whole information about physics of our Universe is focused on studying of this untypical particle physics. In this paper this point of view is touched up on. As the general-relativistic model of our Universe we study the Einstein--Friedmann Spacetime. Firstly, the way of canonical quantization beginning from first quantization of the Dirac Hamiltonian constraints up to the second quantization by the Von Neumann--Araki--Woods quantization in the Fock space is briefly discussed. We show that using of the Bogoliubov--Heinsenberg static operator basis leads to formulation of the second quantization of the considered free boson string in terms of the monodromy in the Fock space. Finally, we propose some specific model of the Universe - \emph{the extremal tachyon mass model}, and in frames of its the Hubble evolution parameter, the equation of state for Dark Matter in the Universe, and the temperature of our Universe are concluded.
\end{abstract}


\maketitle
\section{Introduction}
In my two topical previous papers \cite{g, g1} was discuss in details the so called Many-Particle Quantum Gravity (MPQG) approach to Cosmology, which can be called simply Many-Particle Quantum Cosmology (MPQC). \mbox{I considered} the model of our Universe given by flat, homogenous and isotropic Spacetime historically firstly investigated by Albert Einstein and Alexander Friedmann. In the present paper I want to underlie and discuss completely new context of the proposed approach, that arises from the String Theory. As it is demonstrated in modern literature, this point of view creates beautiful physical picture of the Universe in terms of the fundamental mass groundstate excitation of the free bosonic string - a hypothetical particle with negative mass square and velocity more than the velocity of light $c$, so called \emph{tachyon}. In other words bosonic string theory gives opportunity to understand Cosmology given by the flat Einstein--Friedmann Spacetime in terms of mass groundstate of some wider theory. By this, in the String Theory sense, whole information about physics of our Universe is essentially focused around the studying of dynamics of the tachyon.

This paper is devoted to discuss the string-theory context of the Many-Particle Quantum Cosmology. Firstly, I discuss very concisely primary quantization and secondary one for the Einstein--Friedmann Spacetime with special focusing of attention on monodromy of the Universe as quantum integrable system in the Fock space of creation and annihilation operators. I mention only crucial results of the formal thermodynamics for the tachyon. Except for this old results presented in the new context, I discuss also some interesting new results in the new context. I propose so called \emph{the extremal tachyon mass model} for description of the Universe, that lets determinate unambiguously the Hubble function and the equation of state for the Dark Matter. In frames of this model I compute relation between the Friedmann scale factor and the cosmological time, and I show that the evolution of the Fock space has a standard form. Finally, I compute temperature of the Universe in the extremal tachyon mass model.

\section{Constraints and Strings}
The Standard Cosmology can be understand as studying of the General Relativity formulated by action principle for the Einstein--Hilbert action \cite{h,e1}\footnote{In this paper the units $\hbar=c=k_B=\dfrac{8\pi G}{3}=1$ are used.}
\begin{equation}\label{eh}
\mathit{S}_{\mathrm{EH}}=\int d^4x\sqrt{-g}\left(-\dfrac{1}{6}\mathcal{R}+\mathcal{L}\right),
\end{equation}
where $\mathcal{L}$ is the Lagrangian of the Matter fields, $\mathcal{R}$ is the Ricci curvature, and $g$ is a metric determinant, in case of homogenous, flat, and isotropic Riemannian surface described by the Einstein--Friedmann metric \cite{e,f}
\begin{equation}\label{ef}
ds^2=a^2(\eta)\left[(d\eta)^2-(dx^i)^2\right],~~d\eta=N(x^0)dx^0
\end{equation}
where $\eta$ is conformal time, $x^{\mu}$ $(\mu=0,1,2,3)$ are spacetime coordinates, $a$ is the Friedmann conformal scale factor, and $N$ is the lapse function. In the modern theoretical as well as observational Cosmology the main object of interests and deep studies is this Spacetime with some modifications in form of cosmological perturbations, inflation, and many others (see, \emph{e.g.} \cite{l, b, b1, 0, 1, 2, 3, 4, 5, 6, 7, k}) that we are not going to discuss in this paper. A model of the Universe given by the Spacetime (\ref{ef}) can be understood in terms of the Hamiltonian dynamics of constrained systems proposed by Dirac \cite{d}, and treated in frames metric formalism investigated by Arnowitt, Deser, and Misner \cite{adm}. In result of this approach, the Einstein--Friedmann Universe can be unambiguously characterized by the constraints
\begin{equation}\label{c}
p_{a}^{2}-{\omega}^{2}=0,
\end{equation}
where $p_{a}=-2V_0\dfrac{da}{d\eta}$ is canonical momentum conjugated to the Friedmann scale factor $a$ with finite space volume $V_0=\int d^3x$, and $\omega=\omega(a)$ being in strict relation with energy density of all physical fields in the Universe $\rho(a)$
\begin{equation}
\omega=2V_0\sqrt{\rho(a)},~~\rho(a)=\dfrac{a^4}{V_0}\int{d^{3}x}~\mathcal{H}(x),
\end{equation}
where $\mathcal{H}(x)$ is the Hamiltonian of these fields.

One can see easily that the Dirac Hamiltonian primary constraints (\ref{c}) with formal correspondence
\begin{equation}\label{cor}
m^2 \longrightarrow -\omega^2,
\end{equation}
can be rewritten to form
\begin{equation}
p_a^2+m^2=0,
\end{equation}
that describes the primary constraints for the theory of free bosonic string \cite{lt} with mass $m$. Since formally square of mass for given string is negative, we actually consider the fundamental tachyon excitation that is the mass groundstate of bosonic string theory \cite{gsw,bn,p}. Definition of conformal time $\eta$ proposed by Friedmann uses cosmological time $t$ by relation
\begin{equation}
d\eta=\dfrac{dt}{a(t)},~~t=\tau+x^0,~~\tau=\mathit{constans},
\end{equation}
that together with definition of the Hubble evolution parameter $H(a)$
\begin{equation}
H(a)=\dfrac{1}{a}\dfrac{da}{dt},
\end{equation}
lets do the identification
\begin{equation}
H^2(a)=\dfrac{\rho(a)}{a^4}\equiv\dfrac{1}{V_0}\int{d^{3}x}~\mathcal{H}(x).
\end{equation}
In this manner mass of considered tachyon is directly connected with the Hubble evolution parameter
\begin{equation}\label{mass}
m=m(a)=i2V_0a^2H(a),
\end{equation}
and moreover, the Hubble law, that is a classical solution of primary constraints (\ref{c}), is nontrivial additional condition for the tachyon mass
\begin{equation}
\dfrac{1}{i(t-t_I)}\int_{a_I^2}^{a^2}\dfrac{dy}{m(y)}=\dfrac{1}{V_0},
\end{equation}
where $y=a^2$ is integral variable, and index $I$ means initial data.

Applying to the primary constraints (\ref{c}) the canonical quantization by equal time commutation relations allows introduce conjugate momentum operator $\mathrm{p}_{a}$ according to the standard rules
\begin{equation}\label{4.2}
i\left[\mathrm{p}_{a},a\right]=1~~\Longrightarrow~~\mathrm{p}_{a}=\dfrac{1}{i}\dfrac{\partial}{\partial{a}},
\end{equation}
and in result leaves to the classical field equations in a form
\begin{equation}\label{ham1}
\dfrac{\partial}{\partial{a}}\left[\begin{array}{c}\Psi\\
\Pi_\Psi\end{array}\right]=\left[\begin{array}{cc}
0&1\\
m^{2}&0\end{array}\right]\left[\begin{array}{c}\Psi\\
\Pi_\Psi\end{array}\right],
\end{equation}
where $\Psi$ is the Wheeler--DeWitt wave function \cite{w1,w2}, and \mbox{$\Pi_\Psi$} is classical canonical momentum field conjugated to $\Psi$ understood as classical field. The conception of wave function of the Universe as a solution of the Wheeler--DeWitt equation, being result of compounding of the two equations (\ref{ham1}), was investigated by Hartle and Hawking in \cite{hh} and by Halliwell and Hawking \cite{hah}, but since then this type divagations do not seem clear physical interpretations. By this I propose consider the second quantization of two--component evolution (\ref{ham1}).

\section{Monodromy in the Fock space}
In aim to the second quantization of the theory described by the two-component classical field equations (\ref{ham1}) we must use the quantization that realizes general field--operator canonical commutation relations
\begin{equation}
\left[\mathbf{\Pi_{\Psi}}[a],\mathbf{\Psi}[a']\right]=-i\delta_{aa'},\label{q}
\end{equation}
where $\delta_{aa'}\equiv\delta\left(a-a'\right)$, \mbox{$a\equiv a(\eta)$}, $a'\equiv a(\eta')$ for shortness, together with trivial commutators of two fields $\mathbf{\Psi}$, and two conjugate momenta fields $\mathbf{\Pi_\Psi}$. One can prove easily that the second quantization in a form of the Von Neumann--Araki--Woods quantization \mbox{\cite{ccr1, ccr2}}
\begin{eqnarray}\label{2nd}
\left[\begin{array}{c}\mathbf{\Psi}[a]\\\mathbf{\Pi_\Psi}[a]\end{array}\right]
\!\!=\!\!\left[\begin{array}{cc}\dfrac{1}{\sqrt{2{\omega}}}&\dfrac{1}{\sqrt{2{\omega}}}\\
-i\sqrt{\dfrac{{\omega}}{2}}&i\sqrt{\dfrac{{\omega}}{2}}\end{array}\right]
\left[\begin{array}{c}\mathcal{G}[a]\\ \mathcal{G}^{\dagger}[a]\end{array}\right]\!\!,
\end{eqnarray}
realizes the general relations (\ref{q}) when and only when $\mathcal{G}[a]$ and $\mathcal{G}^{\dagger}[a]$ create bosonic type dynamical functional operator basis $$\mathcal{B}_{a}=\left\{\left[\begin{array}{c}\mathcal{G}[a]\\
\mathcal{G}^{\dagger}[a]\end{array}\right]\!:\!\left[\mathcal{G}[a],\mathcal{G}^{\dagger}[a']\right]\!=\!\delta_{aa'}, \left[\mathcal{G}[a],\mathcal{G}[a']\right]\!=\!0\right\}.$$ Evolution in this basis is governed by quantized classical field theory (\ref{ham1})
\begin{eqnarray}
\dfrac{\partial}{\partial{a}}\left[\begin{array}{c}\mathcal{G}[a]\\
\mathcal{G}^{\dagger}[a]\end{array}\right]=\left[\begin{array}{cc}
-m&\dfrac{1}{2m}\dfrac{\partial m}{\partial a}\\
\dfrac{1}{2m}\dfrac{\partial m}{\partial a}&m\end{array}\right]\left[\begin{array}{c}\mathcal{G}[a]\\
\mathcal{G}^{\dagger}[a]\end{array}\right].\label{fock}
\end{eqnarray}
As it is concluded from detailed analysis \cite{g, g1} that we are not going to present in this paper, the quantum system (\ref{fock}) is fully integrable under very strict and nontrivial conditions. Namely, we must use so called \emph{the Bogoliubov--Heisenberg basis}, that is constructed by applying of the Bogoliubov automorphism in dynamical operator basis $\mathcal{B}_{a}$, and diagonalization of the evolution (\ref{fock}) to the Heisenberg operator equations form in assumed static operator basis $\mathcal{B}_0$ , that is given by standard ladder operators $$\mathcal{B}_{0}=\left\{\left[\begin{array}{c}\mathrm{w}\\
\mathrm{w}^{\dagger}\end{array}\right]: \left[\mathrm{w},\mathrm{w}^{\dagger}\right]=1, [\mathrm{w},\mathrm{w}]=0\right\}.$$ In this basis exists stable quantum vacuum state ad quantum field theory is well-defined. As it can be easily proved the static basis $\mathcal{B}_0$ is related with the dynamical one $\mathcal{B}_{a}$ by the monodromy transformation $\mathbf{M}$ in the Fock space
\begin{equation}\label{mon}
\mathcal{B}_{a}=\mathbf{M}(a)\mathcal{B}_0,
\end{equation}
with the monodromy matrix $\mathrm{\mathbf{M}}(a)$ that equals
\begin{equation}\nonumber
\left[\begin{array}{cc}
\Bigg(\sqrt{\bigg|\dfrac{m}{m_I}\bigg|}+\sqrt{\bigg|\dfrac{m_I}{m}\bigg|}\Bigg)\dfrac{e^{\lambda}}{2}\vspace*{5pt}&
\Bigg(\sqrt{\bigg|\dfrac{m}{m_I}\bigg|}-\sqrt{\bigg|\dfrac{m_I}{m}\bigg|}\Bigg)\dfrac{e^{-\lambda}}{2}\\
\Bigg(\sqrt{\bigg|\dfrac{m}{m_I}\bigg|}-\sqrt{\bigg|\dfrac{m_I}{m}\bigg|}\Bigg)\dfrac{e^{\lambda}}{2}&
\Bigg(\sqrt{\bigg|\dfrac{m}{m_I}\bigg|}+\sqrt{\bigg|\dfrac{m_I}{m}\bigg|}\Bigg)\dfrac{e^{-\lambda}}{2}\end{array}\right],
\end{equation}
where $\lambda$ is integrated mass of considered free bosonic string
\begin{eqnarray}\label{im}
\lambda=\lambda(a)=\pm\int_{a_I}^{a}m~da.
\end{eqnarray}
The whole information about an initial data is contained in this monodromy matrix. By this the quantum cosmology given by the static basis $\mathcal{B}_0$ has the stable vacuum state independent on an initial data. In frames of this well-defined quantum theory we can build formal thermodynamics for the considered mass groundstate.
\section{The Tachyon Thermodynamics}
In frames of presented quantum field theory with stable vacuum state we can build formal thermodynamics of the Universe modeled by tachyon. The fastest method is using of the density matrix formulation. To calculate the formal von Neumann--Boltzmann entropy of discussed system we use the one-particle density functional given by occupation number operator in the dynamical operator basis $\mathcal{B}_{a}$ transformed to the static one $\mathcal{B}_0$. In the static basis, which has stable vacuum and by this thermodynamics is equilibrated, we compute occupation number of quantum states created from the stable vacuum state. By applying of the Gibbs ensemble one can compute formally internal energy and chemical potential for the considered system. Direct identification of calculated partition function with the Bose--Einstein statistics, that is agreeable with conception of tachyon as groundstate of bosonic string, gives in conclusion a relation between the temperature and the Friedmann scale factor $a$ that is a degree of freedom of the considered system. In this manner in frames of MPQC one can obtain formal description of thermodynamical properties of the Universe. Basic relations are presented below
\paragraph{Occupation number $\mathrm{n}=\mathrm{n}(a)$}
\begin{equation}\label{n}
\mathrm{n}=\dfrac{1}{4}\left|\sqrt{\left|\dfrac{m}{m_I}\right|}-\sqrt{\bigg|\dfrac{m_I}{m}\bigg|}\right|^2,~~\langle\mathrm{n}\rangle=2\mathrm{n}+1
\end{equation}
\paragraph{Entropy $\mathrm{S}=\mathrm{S}(a)$}
\begin{equation}\label{ent}
\mathrm{S}=-\ln(2\mathrm{n}+1).
\end{equation}
\paragraph{Internal energy $\mathrm{U}=\mathrm{U}(a)$}
\begin{equation}\label{U}
\mathrm{U}=\left(\dfrac{1}{2}+\dfrac{4\mathrm{n}+3}{2\mathrm{n}+1}\mathrm{n}\right)|m|.
\end{equation}
\paragraph{Chemical potential $\mu=\mu(a)$}
\begin{equation}\label{mu}
\mu=\Bigg(1+\dfrac{1}{(2\mathrm{n}+1)^2}-\dfrac{1}{2}\dfrac{4\mathrm{n}+1}{4\mathrm{n}^2+2\mathrm{n}}\sqrt{\dfrac{\mathrm{n}}{\mathrm{n}+1}}\Bigg)|m|.
\end{equation}
\paragraph{Temperature $\mathrm{T}=\mathrm{T}(a)$}
\begin{eqnarray}\label{temp}
\mathrm{T}=\dfrac{1+\left(\dfrac{2\mathrm{n}}{2\mathrm{n}+1}\right)^2+\dfrac{8\mathrm{n}^2+8\mathrm{n}+1}{4\mathrm{n}+2}\sqrt{\dfrac{\mathrm{n}}{\mathrm{n}+1}}}{2\ln(2\mathrm{n}+2)}|m|.
\end{eqnarray}

\section{The extremal tachyon mass model}
In this section we will discuss some very special case of presented formalism - \emph{the extremal tachyon mass model}.
\subsection{Extremal Hubble parameter}
Now we concentrate our considerations on the integrated mass of the tachyon (\ref{im})
\begin{equation}
\lambda=\lambda(a)=\int_{a_I}^a m(a)da.
\end{equation}
One can see from the monodromy matrix (\ref{mon}) that $\lambda$ can be understood as the Weyl characteristic scale for the Universe. Formally it means that the scale must fulfill the d'Alembert equation
\begin{equation}
\Delta\lambda=0,
\end{equation}
that in considered 1-dimensional case means
\begin{equation}\label{sc}
\dfrac{\partial^2\lambda(a)}{\partial a^2}=0,
\end{equation}
or in terms of the tachyon mass
\begin{equation}\label{mm}
\dfrac{\partial m(a)}{\partial a}=0\Rightarrow m(a)=m(a_I)\equiv m_I.
\end{equation}
Direct using of the definition (\ref{mass}) products the conclusion that the Hubble evolution parameter for the tachyon must has the form as follows
\begin{equation}\label{hue}
H(a)=\dfrac{Q}{a^2},
\end{equation}
where $Q=\dfrac{m_I}{2iV_0}=a_I^2H(a_I)$ is a constant. By the extremal condition (\ref{mm}) for the tachyon mass, we propose call name the Hubble evolution parameter (\ref{hue}) as \emph{the extremal Hubble parameter}.
\subsection{Equation of State for Dark Matter}
In current literature devoted to Cosmology (see, \emph{e.g.} \cite{k, m}) the most often used is the standard cosmological model given by flat, homogenous, and anisotropic the Einstein--Friedmann Universe for that presence of physical fields is modeled by the Hubble evolution parameter
\begin{eqnarray}\label{kh}
H(a)=H_0\sqrt{\epsilon_T+\epsilon_M+\epsilon_R+\epsilon_w},
\end{eqnarray}
where $H_0$ is present-day value of the Hubble parameter,
\begin{eqnarray}
\epsilon_T&=&(1-\Omega_{T})\left(\dfrac{a_I}{a}\right)^2,\\
\epsilon_M&=&\Omega_M\left(\dfrac{a_I}{a}\right)^3,\\
\epsilon_R&=&\Omega_R\left(\dfrac{a_I}{a}\right)^4,\\
\epsilon_w&=&\Omega_w\left(\dfrac{a_I}{a}\right)^3\exp\left\{-3\int_{a_I}^{a}da\dfrac{w(a)}{a}\right\},\label{dm}\\
\Omega_T&=&\Omega_M+\Omega_R+\Omega_w,
\end{eqnarray}
$\Omega$s are density energy of Matter ($M$), Radiation ($R$), Dark Matter ($w$), and ($\Omega_T$) is a total density energy of physical fields in the Universe. The term with a coefficient $1-\Omega_{T}$ is famous as curvature term. In (\ref{dm}) $w(a)=\dfrac{p(a)}{\rho(a)}$ ia equation of state for the Dark Matter. From physical point of view the total $\epsilon_T$ and the radiative $\epsilon_R$ energies give contribution to the Cosmic Microwave Background radiation, the Matter term $\epsilon_M$ is responsible for the Large Scale Structure of the Universe, and the Dark Matter part $\epsilon_w$ describes dark energy contribution. Comparison of the Hubble evolution parameter in the form (\ref{kh}) with the extremal one (\ref{hue}) allow determinate unambiguously initial data $H(a_I)$
\begin{eqnarray}
H(a_I)=H_0\sqrt{\Omega_R}.
\end{eqnarray}
Furthermore, the extremal tachyon mass model leads to the equation of state for Dark Matter
\begin{equation}\label{es}
w(a)=\left\{\begin{array}{cc}\dfrac{1}{3}\sum_{n=0}^{\infty}\left(\dfrac{a}{a_{cr}}\right)^{n+1}&,~\mathrm{for}~a<a_{cr}\vspace*{10pt}\\
-\dfrac{1}{3}\sum_{n=0}^{\infty}\bigg(\dfrac{a_{cr}}{a}\bigg)^{n}&,~\mathrm{for}~a>a_{cr}\end{array}\right.
\end{equation}
where $a_{cr}=\dfrac{\Omega_M}{\Omega_T-1}a_I$ is critical value of $a$, for that the equation of state (\ref{es}) has simple singularity.
\subsection{Extremal thermodynamics}
One can see easily that direct solution of the Dirac constraints for the extremal tachyon mass model (\ref{hue}) has a form
\begin{equation}
a(t)=a_I\sqrt{1+2H_0\sqrt{\Omega_R}|t-t_I|},
\end{equation}
and in result the dependence on cosmological time for the scale $\lambda$ is
\begin{equation}
\lambda(t)=m_Ia_I\left|\sqrt{1+2H_0\sqrt{\Omega_R}|t-t_I|}-1\right|.
\end{equation}
For considered model, the operator evolution (\ref{mon}) has a simple form
\begin{equation}\nonumber
\left[\begin{array}{c}\mathcal{G}(t)\\
\mathcal{G}^{\dagger}(t)\end{array}\right]=\left[\begin{array}{cc}e^{-\lambda(t)}&0\\0&e^{\lambda(t)}\end{array}\right]\left[\begin{array}{cc}\mathrm{w}\\
\mathrm{w}^{\dagger}\end{array}\right].
\end{equation}
Moreover, is consequence of the relation (\ref{mm}), which means that the occupation number (\ref{n}) of quantum states produced from the stable vacuum state is trivially equal to $0$ (or in other words, averaged occupation number is equal to $1$), one can obtain in result the value of the temperature of the Universe as tachyon (\ref{temp}) as
\begin{equation}\label{temp1}
\mathrm{T}=\dfrac{|m_I|}{2\ln2}=V_0\dfrac{a_I^2H_0\sqrt{\Omega_R}}{\ln2}.
\end{equation}
It is amazing that, in spite of finite and constant value of temperature (\ref{temp1}), value of formal Von Neumann-Boltzmann entropy (\ref{ent}) is trivially equal to $0$, but internal energy (\ref{U}) is finite too and equals $\mathrm{U}=\dfrac{1}{2}|m_I|$. It can be seen that the chemical potential (\ref{mu}), as order derivative of an internal energy with expect to occupation number, in the extremal tachyon mass model has a finite and constant value
\begin{equation}
\mu(|m|=|m_I|)= 3|m_I|.
\end{equation}
From the open system theory point of view it means simply that the extremal tachyon mass model describes the Universe as open quantum system with stabilized number of quantum states produced from vacuum state. The extremal tachyon mass model has a beautiful physical meaning -it is the minimal model for the theory given by the monodromy matrix $\mathbf{M}(a)$ in (\ref{mon}). It means that all more general models can be obtained by considering of corrections to the extremal mass model. It is evident from the form of the extremal Hubble parameter (\ref{hue}) that the minimal model means presence of the Radiation only, and more general models are studying of the Matter and the Dark Matter contributions. 
\section{Many-Particle Quantum Gravity}
What is the Many-Particle Quantum Cosmology truly? In the present paper I have showed the String Theory face of this approach, but it can be formulate equivalently also in language of the open system theory. Truly, presented approach formulates the Quantum Cosmology by dynamics of fundamental tachyon excitation of free bosonic string, and it is \emph{the correct} definition of the Many-Particle Quantum Cosmology.

What is a crucial problem for correct formulation of Quantum Gravity? Serving the analogy method to presented above approach, one can say that the problem lies in investigating of bosonic string dynamics and formulation of the second quantization of first-quantized theory.

The author nurture hopes that Many-Particle Quantum Gravity approach will find further applications.
\section*{Acknowledgements}

The author benefitted from many valuable discussions, notably with \mbox{B.M. Barbashov}, \mbox{I. Bia\l ynicki-Birula}, \mbox{S.J. Brodsky}, \mbox{S. Ferrara}, \mbox{G. $\acute{}$ t Hooft}, \mbox{L.N. Lipatov}, \mbox{S. Pokorski},  and \mbox{D.V. Shirkov}.

\end{document}